

%
\magnification=\magstep1        
\font\tenbm=cmmib10		
\newfam\bmfam
\textfont\bmfam=\tenbm
\def\bmit{\fam\bmfam\tenbm}
\mathchardef\Theta="7102
\mathchardef\Phi="7108
\mathchardef\delta="710E
\font\tenmsb=msbm10		
\newfam\msbfam
\textfont\msbfam=\tenmsb
\catcode`\@=11
\def\hexnumber@#1{\ifcase#1 0\or 1\or 2\or 3\or 4\or 5\or 6\or 7\or 8\or
 9\or A\or B\or C\or D\or E\or F\fi}
\edef\msbfam@{\hexnumber@\msbfam}
\def\Bbb#1{\fam\msbfam\relax#1}
\catcode`\@=12

%
\font\bigrm=cmr10 scaled 1200		
\font\bigbf=cmbx10 scaled 1200		
\def\bigbold{\large\bf}

\def\large{\def\rm{\fam0\bigrm}\textfont\bffam=\bigbf	
 \def\bf{\fam\bffam\bigbf}\rm}
\def\normal{\def\rm{\fam0\tenrm}\textfont\bffam=\tenbf	
 \def\bf{\fam\bffam\tenbf}\rm}
\catcode`\@=11				
\def\footnote#1{\edef\@sf{\spacefactor\the\spacefactor}#1\@sf
 \insert\footins\bgroup\normal \interlinepenalty100 \let\par=\endgraf
 \leftskip=0pt \rightskip=0pt \splittopskip=10pt plus 1pt minus 1pt
 \floatingpenalty=20000
 \smallskip\item{#1}\bgroup\strut\aftergroup\@foot\let\next}
\catcode`\@=12

\newcount\EQNO      \EQNO=0
\newcount\FIGNO     \FIGNO=0
\newcount\REFNO     \REFNO=0
\newcount\SECNO     \SECNO=0
\newcount\SUBSECNO  \SUBSECNO=0
\newcount\FOOTNO    \FOOTNO=0
\newbox\FIGBOX      \setbox\FIGBOX=\vbox{}
\newbox\REFBOX      \setbox\REFBOX=\vbox{}
\newbox\RefBoxOne   \setbox\RefBoxOne=\vbox{}

\expandafter\ifx\csname normal\endcsname\relax\def\normal{\null}\fi

\def\Eqno{\global\advance\EQNO by 1 \eqno(\the\EQNO)%
    \gdef\label##1{\xdef##1{\nobreak(\the\EQNO)}}}
\def\Fig#1{\global\advance\FIGNO by 1 Figure~\the\FIGNO%
    \global\setbox\FIGBOX=\vbox{\unvcopy\FIGBOX
      \narrower\smallskip\item{\bf Figure \the\FIGNO~~}#1}}
\def\Ref#1{\global\advance\REFNO by 1 \nobreak[\the\REFNO]%
    \global\setbox\REFBOX=\vbox{\unvcopy\REFBOX\normal
      \smallskip\item{\the\REFNO .~}#1}%
    \gdef\label##1{\xdef##1{\nobreak[\the\REFNO]}}}
\def\Section#1{\SUBSECNO=0\advance\SECNO by 1
    \bigskip\leftline{\bf \the\SECNO .\ #1}\nobreak}
\def\Subsection#1{\advance\SUBSECNO by 1
    \medskip\leftline{\bf \ifcase\SUBSECNO\or
    a\or b\or c\or d\or e\or f\or g\or h\or i\or j\or k\or l\or m\or n\fi
    )\ #1}\nobreak}
\def\Footnote#1{\global\advance\FOOTNO by 1
    \footnote{\nobreak$\>\!{}^{\the\FOOTNO}\>\!$}{#1}}
\def\SameFootnote{$\>\!{}^{\the\FOOTNO}\>\!$}

\def\References{\bigskip\centerline{\bf REFERENCES}
                \smallskip\copy\REFBOX}
\def\NewRefPage{\setbox\RefBoxOne=\vbox{\unvcopy\REFBOX}%
		\setbox\REFBOX=\vbox{}%
		\def\References{\bigskip\centerline{\bf REFERENCES}
                		\nobreak\smallskip\nobreak\copy\RefBoxOne
				\vfill\eject
				\smallskip\copy\REFBOX}%
		\def\NewRefPage{}}


\def\MultiRef#1{\global\advance\REFNO by 1 \nobreak\the\REFNO%
    \global\setbox\REFBOX=\vbox{\unvcopy\REFBOX\normal
      \smallskip\item{\the\REFNO .~}#1}%
    \gdef\label##1{\xdef##1{\nobreak[\the\REFNO]}}}

\def\sgn{{\rm sgn}}
\def\AtSigma{|_{\scriptscriptstyle\Sigma}}

\def\Lim#1#2{\!\!\!\lim_{~~t\to#1^#2}\!}
\def\LongInt#1#2#3{\hskip 1em #1
   \hskip 1em {#2} \hskip -1em
   \int\limits_{\rm #3} \hskip -1em}
\def\Half{\hbox{$\displaystyle 1 \over \displaystyle 2$}}

\def\p{\partial}

\def\upm{{U^{\pm}}}
\def\upms{{\scriptscriptstyle\upm}}
\def\ypm{{\bmit\Theta^{\pm}}}
\def\smh{\sqrt{\vert t\vert}}
\def\d{{\bmit\delta}}
\def\fp{F^+\AtSigma}
\def\fm{F^-\AtSigma}
\def\s{\{t=0\}}
\def\bff{{\bmit F}}
\def\bfy{{\bmit\Theta}}

\overfullrule=0pt		

\def\today{\number\day\space\ifcase\month\or
  January\or February\or March\or April\or May\or June\or
  July\or August\or September\or October\or November\or December\fi
  \space\number\year}
\rightline{25 January 1995}
\rightline{gr-qc/9501034}
\bigskip\bigskip


\null\bigskip
\centerline{\bigbold BOUNDARY CONDITIONS FOR THE SCALAR FIELD}
\smallskip
\centerline{\bigbold IN THE PRESENCE OF SIGNATURE CHANGE }
\bigskip

\centerline{Tevian Dray
\Footnote{Permanent address is Oregon State University.}
}
\centerline{\it School of Physics \& Chemistry, Lancaster University,
		Lancaster LA1 4YB, UK}
\centerline{\it Department of Mathematics, Oregon State University,
		Corvallis, OR  97331, USA}
\centerline{\tt tevian{\rm @}math.orst.edu}
\medskip
\centerline{Corinne A. Manogue
\SameFootnote
}
\centerline{\it School of Physics \& Chemistry, Lancaster University,
		Lancaster LA1 4YB, UK}
\centerline{\it Department of Physics, Oregon State University,
		Corvallis, OR  97331, USA}
\centerline{\tt corinne{\rm @}physics.orst.edu}
\medskip
\centerline{Robin W. Tucker}
\centerline{\it School of Physics \& Chemistry, Lancaster University,
		Lancaster LA1 4YB, UK}
\centerline{\tt R.W.Tucker{\rm @}lancaster.ac.uk}

\bigskip\bigskip
\centerline{\bf ABSTRACT}
\midinsert
\narrower\narrower\noindent
We show that, contrary to recent criticism, our previous work yields a
reasonable class of solutions for the massless scalar field in the presence of
signature change.
\endinsert
\bigskip\bigskip

\Section{INTRODUCTION}

In a recent letter
\Ref{Sean A. Hayward, {\it Weak Solutions Across a Change of Signature},
Class.\ Quantum Grav.\ {\bf 11}, L87 (1994).}\label\Hayward
\xdef\HAYWARD{\the\REFNO}%
, Hayward purports to show that our earlier work on signature change,
especially
\Ref{Tevian Dray, Corinne A. Manogue, and Robin W. Tucker,
{\it Particle Production from Signature Change},
Gen.\ Rel.\ Grav.\ {\bf 23}, 967 (1991).} \label\PaperI
\xdef\PAPERI{\the\REFNO}%
but also
\Ref{Tevian Dray, Corinne A. Manogue, and Robin W. Tucker,
{\it The Scalar Field Equation in the Presence of Signature Change},
Phys.\ Rev.\ {\bf D48}, 2587 (1993).}\label\PaperIII
\xdef\PAPERIII{\the\REFNO}%
, is ``mathematically inconsistent'', ``entails a non-uniqueness which
destroys predictability'', and does not ``make sense of the relevant
field equations''.  He has made similar criticisms elsewhere
\Ref{Sean A. Hayward,
{\it Junction conditions for signature change},
gr-qc/9303034.}\label\HaywardII
.  Contrary to Hayward's claims, our approach is
completely consistent and makes sense of our field equation.

In many situations in physics one deals with equations that admit singular
solutions.  Sometimes one can simply excise those domains of the manifold
where the singularity occurs and replace the effect of the singularity by a
suitable set of parameters.  But sometimes the components of tensor equations
themselves are singular.  In this case a strategy must be adopted to formulate
the problem before attempting a solution.  This is precisely the situation
under discussion here, where a scalar field equation is sought on a manifold
with a degenerate metric which changes signature at a hypersurface.  Our
formulation of the problem is different from Hayward's, leading to different
equations with different solution spaces.  In the absence of experimental
guidance, both approaches are viable.

Hayward's approach
[\HAYWARD,%
\MultiRef{Sean A. Hayward,
Class.\ Quantum Grav. {\bf 9}, 1851 (1992);
\hfill\break
erratum: Class.\ Quantum Grav. {\bf 9}, 2543 (1992).}]
is to give a global definition of the scalar field equation on such a manifold
and to demand that, since the field equation is second order, its solutions be
globally $C^2$.  He concludes that the field momentum must vanish at the
surface of signature change.  Our approach \PaperIII\ is to demand instead
that the scalar field equation be defined in a piecewise fashion on the
manifold and to admit piecewise smooth solutions.  We then choose conditions
that match the piecewise solutions across the degeneracy hypersurface.  Our
conditions are simply that the field and the unit normal derivative of the
field (or equivalently the canonical momentum) exist and be continuous.  These
conditions can be derived by promoting the piecewise formulation of the field
equation into a single global {\it distributional} equation \PaperIII.
Hayward \Hayward\ dismisses our approach claiming that it leads to non-unique
solutions and hence ``destroys predictability''.  This claim is false.  We
show below how solutions may be completely determined up to normalization.
Our procedure does however select a class of solutions that are not
necessarily globally $C^2$.

We note with interest that recent work
[\MultiRef{L. J. Alty,
{\it Kleinian Signature Change},
Class.\ Quantum Grav.\ {\bf 11}, 2523 (1994).},%
\MultiRef{L. J. Alty and C. J. Fewster,
{\it Initial Value Problems and Signature Change},
DAMTP preprint R95/1, gr-qc/9501026 (1995).}]
in the context of Kleinian signature change also argues in favor of a
continuity condition on the unit normal derivative of the field, rather than
requiring that it vanish.

We begin this rebuttal by giving the simple example, first introduced in
\PaperI, which Hayward uses in his attempt \Hayward\ to show that our
solutions are non-unique.  We demonstrate that our approach does indeed
determine the arbitrary constants.

In Section 3, we examine several different generalizations of the standard
action for the massless scalar field to a signature changing background.  For
each such generalization, a variational principle requires that the
appropriate canonical momentum be continuous (and in one case, zero) at the
boundary.  In this way, we see how both our boundary conditions and Hayward's
can be derived in a parallel manner, but from different actions.  From this
point of view, it is not surprising that the theories obtained from these
actions have different spaces of classical solutions.  We also discuss the
implications of each of the resulting theories.

Contrary to Hayward's claims, our approach can be derived from an explicit
field equation using standard techniques.  We discuss our field equation in
Section 4.  The fact that our distributional field equation is not the one
that Hayward uses seems to have been lost in his claims that we do not make
sense of ``the'' field equations.  To discuss a scalar field equation on a
manifold with a singular metric, one must first formulate the problem in an
unambiguous manner.  It is simply incorrect to refer to ``the'' scalar field
equation on such a manifold.  In the Appendix, we further show that our
approach yields the standard junction conditions for discontinuous Maxwell
fields.

An important aspect of quantum cosmology concerns semiclassical approximations
to path integrals, which exploit particular classical solutions --- hence the
desire of some to look at ``real-tunneling'' solutions.  Our theory here is
purely classical, and we wish to consider all classical solutions.  We discuss
these differing motivations in Section 5.

\Section{EXAMPLE}

The differences in the two approaches can be best seen in terms of an example,
first introduced in \PaperI\ and also discussed in \Hayward.  Consider the
singular differential equation
$$
2t\,\ddot\Phi = \dot\Phi \Eqno
$$\label\ODE
This equation can be viewed as the  massless scalar wave equation
for the 1-dimensional signature-changing metric $t\,dt^2$.

The general solution to \ODE\ is
$$
\Phi = \Bigg\{
		\matrix{A (-t)^{3/2} + B & (t<0)\cr
			\noalign{\vskip5pt}
			C t^{3/2} + D & (t>0)}
  \Eqno
$$ \label\Sol
where $A$, $B$, $C$, $D$ are constants.  The requirement that $\Phi$ be
continuous at $t=0$ fixes $D=B$.

\goodbreak
Our approach \PaperIII\ is to demand in addition that the (unit) normal
derivative of $\Phi$ at $t=0$, defined in terms of 1-sided limits, be
continuous, namely
$$
  \Lim0+ {~\dot\Phi\over\sqrt{t}} = \Lim0- {~\dot\Phi\over\sqrt{-t}}\Eqno
$$ \label\Match
This fixes $C=-A$ in \Sol, and the solution becomes
\Footnote{As discussed in \PaperIII, an alternative boundary condition,
corresponding to the freedom in choosing the relative orientation of the
normal derivatives, is to insert a minus sign on one side of \Match, which
would lead to $C=A$ in \Sol.  Both of these (classes of) solutions were given
in \PaperI.}
$$
\Phi = \Bigg\{
		\matrix{A (-t)^{3/2} + B & (t<0)\cr
			\noalign{\vskip5pt}
			-A t^{3/2} + B & (t>0)}
  \Eqno
$$
The integration constant $B$ can be determined, e.g.\ by the choice of
$\Phi(0)$, and $A$ can be fixed by appropriate normalization.  If however,
following Hayward \Hayward, one instead demands that $\Phi$ be globally $C^2$,
then the only solution is $\Phi=B$.  Each choice leads to its own particular
class of solutions.

Alternatively, in \Hayward, Hayward examines the solutions \Sol\ without
imposing any further conditions such as \Match.  It is these solutions which
Hayward claims exhibit non-uniqueness.  We have never suggested using
solutions of this form.

The controversy, therefore, apparently comes down to whether or not it is
reasonable to allow solutions to a second order differential equation which
are not globally $C^2$.  This is commonly done in elementary physics across
regions which contain physical discontinuities such as the electrostatic field
across a hollow charged conductor; see the Appendix.

\Section{VARIATIONAL APPROACH}

The standard action for a real massless scalar field on an $n$-dimensional
(non-degen\-erate) Lorentzian background is:
$$
{\cal S}_L = {\Half}\int g^{\mu \nu} \Phi,_{\mu} \Phi,_{\nu}
		\sqrt{-g}\,d^{n-1} x \, dt
\Eqno$$\label\LAction
There are several possible generalizations of this action which might be used
to accommodate a background spacetime which changes signature from Lorentzian
to Euclidean at a hypersurface $\Sigma=\{t={\rm constant}\}$.  It is
illuminating to examine the consequences of varying these actions.

One possibility is to extend \LAction\ to the Euclidean region by putting an
absolute value in the square root.  Such an action has the advantage that it
naturally remains real even in the Euclidean region:
$$
{\cal S}_1 \LongInt={\Half}{Lorentzian}
  g^{\mu \nu} \Phi,_{\mu} \Phi,_{\nu} \sqrt{\vert g\vert}\, d^{n-1} x \, dt
  \LongInt+{\Half}{Euclidean}
  g^{\mu \nu} \Phi,_{\mu} \Phi,_{\nu} \sqrt{\vert g\vert}\, d^{n-1} x \, dt
\Eqno$$\label\Sone
We vary ${\cal S}_1$ and demand that it be stationary for variations of $\Phi$
that
do not necessarily vanish on $\Sigma$.  The variation is
$$
\eqalign{\delta {\cal S}_1 \LongInt={-}{Lorentzian} &
	\left(g^{\mu \nu} \Phi,_{\mu} \sqrt{\vert g\vert}\right),_{\nu} \,
	\delta \Phi\, d^{n-1} x \, dt
{}~~+~~ \int\limits_\Sigma
	\left(g^{0 \nu} \Phi,_{\nu}
	\sqrt{\vert g\vert}\right) \, \delta\Phi\, d^{n-1}x \cr
\LongInt{}{-}{Euclidean~} &
	\left(g^{\mu \nu} \Phi,_{\mu} \sqrt{\vert g\vert}\right),_{\nu} \,
	\delta \Phi\, d^{n-1} x \, dt
{}~~-~~ \int\limits_\Sigma
	\left(g^{0 \nu} \Phi,_{\nu}
	\sqrt{\vert g\vert}\right) \, \delta\Phi\, d^{n-1}x \cr}
\Eqno$$
and we immediately obtain the scalar field equations in the Lorentzian and
Euclidean regions separately from variations whose support does not intersect
$\Sigma$.  If we now seek solutions such that $\Phi$ is continuous, it is
natural to assume that $\delta\Phi$ is continuous at $\Sigma$, and the
remaining variations yield a boundary condition on the canonical momentum
\Footnote{For all of the actions in this section, we consider only fields for
which the canonical momenta have bounded limits to $\Sigma$; this implies that
the integrands of the various actions are well-behaved near $\Sigma$.  The
variation of each action then yields the usual scalar field equations in the
Lorentzian and Euclidean regions separately; it is the boundary conditions
which differ in each case.}
$$
\Pi_1 := {\delta {\cal S}_1\over \delta \Phi,_0} =
         \sqrt{\vert g\vert}\, g^{0\nu}\, \Phi,_{\nu}
\Eqno$$\label\Pione
namely that $\Pi_1$ be continuous at $\Sigma$.  This is precisely the boundary
condition which we have proposed [\PAPERI,\PAPERIII]; it is equivalent to
continuity of the normal derivative, as in \Match.  Note that this boundary
condition is required by a variational principle.  Solutions of the equations
of motion for which $\Phi$ and $\Pi_1$ are continuous do not suffer from any
``non-uniqueness'' which ``destroys predictability''.  Quite the contrary,
{\bf any} real Euclidean solution is matched to precisely one Lorentzian
solution by this prescription.

We showed in \PaperIII\ that the Klein-Gordon product of such solutions is
conserved even across the hypersurface of signature change.  This is a
necessary condition for a corresponding quantum theory to be unitary and is
thus an attractive feature of this approach.

The action ${\cal S}_1$ can be extended to the complex scalar field in the
usual way, yielding the action
$$
{\cal S}_2 \LongInt={}{Lorentzian}
  g^{\mu \nu} \Phi,_{\mu} \Phi^*,_{\nu} \sqrt{\vert g\vert}\, d^{n-1} x \, dt
  \LongInt+{}{Euclidean}
  g^{\mu \nu} \Phi,_{\mu} \Phi^*,_{\nu} \sqrt{\vert g\vert}\, d^{n-1} x \, dt
\Eqno$$\label\Sone
Because ${\cal S}_2$ is real, it is essentially two copies of the action for
the real scalar field, as usual.  The boundary condition obtained from varying
${\cal S}_2$ is just the continuity of the (now complex) canonical momentum
\Pione.  Such an action is a natural candidate for extension, via minimal
coupling to an electromagnetic field in a $U(1)$ invariant way.

A third possibility involves extending \LAction\ to the Euclidean region via
analytic continuation of $\sqrt{-g}$ in the variable $t$, yielding
$$
{\cal S}_3 \LongInt={}{Lorentzian}
  g^{\mu \nu} \Phi,_{\mu} \Phi,_{\nu} \sqrt{-g}\, d^{n-1} x \, dt
  \LongInt+{}{Euclidean}
  g^{\mu \nu} \Phi,_{\mu} \Phi,_{\nu} \sqrt{-g}\, d^{n-1} x \, dt
\Eqno$$\label\Sthree
where we have implicitly chosen a branch.  We allow $\Phi$ to take complex
values, even in the Lorentzian region, and discuss later the consequences of
restricting to real $\Phi$.

Varying the real part of ${\cal S}_3$ yields the boundary condition that the
appropriate canonical momentum, given by
$$
\Pi_3 := {\delta {\cal S}_3\over \delta \Phi,_0} =
         \sqrt{-g}\, g^{0\nu}\, \Phi,_{\nu}\Eqno
$$\label\Pithree
should be continuous at $\Sigma$.  Notice that this canonical momentum differs
from the previous one \Pione\ by a factor of $i$ in the Euclidean region.  The
imaginary part of ${\cal S}_3$ is redundant, yielding the same field equations
and the same boundary condition, \Pithree, so that the real part of
${\cal S}_3$ can be used alone as a real action for this theory.

Unlike ${\cal S}_1$ and ${\cal S}_2$, the Klein-Gordon products of some
solutions of the equations of motion matched by \Pithree\ are not conserved
across the hypersurface of signature change, so that it would be difficult to
build a unitary quantum theory from such an action.  Also, it is not obvious
how to couple the scalar field in \Sthree\ to an electromagnetic field in a
$U(1)$ invariant way, even in the Lorentzian region.

If one makes the added restriction that solutions $\Phi$ must be real in both
the Lorentzian and Euclidean regions, it is impossible to satisfy continuity
of $\Pi_3$ unless the canonical momentum is zero at the boundary.  This is the
junction condition which Hayward prefers.  Solutions of this type do have
conserved Klein-Gordon product, but the class of allowable solutions is
restricted by this extra reality condition.  We discuss in Section 5 why such
{\it particular} solutions are of relevance in the context of a Euclidean
approach to the quantum theory.

A fourth possibility involves starting with the conventional action for the
complex massless scalar field in the Lorentzian region and using the analytic
extension of $\sqrt{-g}$ into the Euclidean region.  Then the action becomes
$$
{\cal S}_4 \LongInt={}{Lorentzian}
  g^{\mu \nu} \Phi,_{\mu} \Phi^*,_{\nu} \sqrt{-g}\, d^{n-1} x \, dt
  \LongInt+{}{Euclidean}
  g^{\mu \nu} \Phi,_{\mu} \Phi^*,_{\nu} \sqrt{-g}\, d^{n-1} x \, dt
\Eqno$$\label\Sfour
This results in a non-analytic integrand and a complex action.  By varying
${\cal S}_4$ with respect to the real and imaginary parts of $\Phi$ we obtain
the equations of motion together with the following boundary conditions:
$$
\eqalign{\Lim\Sigma- \sqrt{-g}\, g^{0\nu}\, \Phi,_{\nu}
       &=\Lim\Sigma+ \sqrt{-g}\, g^{0\nu}\, \Phi,_{\nu}\cr
         \Lim\Sigma- \sqrt{-g}\, g^{0\nu}\, \Phi,_{\nu}^*
       &=\Lim\Sigma+ \sqrt{-g}\, g^{0\nu}\, \Phi,_{\nu}^*\cr}
\Eqno$$\label\Haymatch

This action is complex, so that the boundary condition on $\Phi^*$ is not the
conjugate of the condition on $\Phi$, yielding an additional constraint on the
allowed solutions.  We find this property of complex actions like ${\cal S}_4$
rather disturbing, as it restricts the allowed solutions away from the
boundary in a way which the other actions considered here do not.  Solving
both equations in \Haymatch\ simultaneously requires both sides of each
equation to be zero; this is again the junction condition which Hayward
prefers.  However, as the resulting solution space turns out to be a proper
subset of that determined by ${\cal S}_1$, it is clear that Klein-Gordon
products are preserved.  Furthermore, and despite the action being complex,
this theory is a candidate for extension, via minimal coupling to an
electromagnetic field in a $U(1)$ invariant way.

Of the two real actions considered here, $S_1$ and the real part of $S_3$,
only the boundary condition obtained from a variational principle for $S_1$
matches real solutions to real solutions.  Furthermore, only $S_1$ conserves
Klein-Gordon products of solutions across the signature change unless the
solution space for $S_3$ is additionally restricted by a reality condition.
Of the actions for the complex field considered here, both $S_2$ and $S_4$
conserve Klein-Gordon products of solutions and both appear to provide
candidates for minimal coupling to an electromagnetic field.  However, in the
case of $S_4$ our procedure yields two boundary conditions, instead of the
expected one, since the boundary condition on $\Phi^*$ is not the complex
conjugate of the boundary condition on $\Phi$.

The set of actions given here is by no means exhaustive.  There are clearly
many other possibilities, such as other relative factors between the
Lorentzian and Euclidean actions, or the addition of surface terms on
$\Sigma$, leading to discontinuous canonical momenta.  Nevertheless, we feel
that this language is a good one for illustrating how a variety of boundary
conditions can be obtained.  It also provides a framework for discussing some
of the considerations which might arise when one attempts to find a physical
interpretation of the extension of the theory of the massless scalar field to
a signature changing background.  A recent paper by Embacher
\Ref{Franz Embacher,
{\it Actions for signature change},
University of Vienna preprint UWThPh-1995-1, gr-qc/9501004 (1995).}
\label\Embacher
considers similar generalizations of the Einstein-Hilbert action to
accommodate signature change.

\Section{DISTRIBUTIONAL WAVE EQUATION}

An alternate approach is to use the language of tensor distributions.  We
summarize here our approach to the massless scalar field in this language, as
presented in detail in \PaperIII, and give a 1-dimensional example.

Suppose one is given two regions $U^\pm$ of a manifold $M$ sharing a common
boundary $\Sigma$ given by $\{\chi=0\}$.  Suppose further that the field
equations $dF^\pm=0$ hold separately on the 2 regions, where $F^{\pm}$ are
differential forms on $U^\pm$.  Introduce the distributional field
$$
\bff = \bfy^+ F^+ + \bfy^- F^-\Eqno
$$\label\Dist
in terms of the Heaviside distributions $\ypm$ with support in $\upm$ and such
that
$$
d\ypm=\pm\d \Eqno
$$
where $\d=\delta(\chi)\,d\chi$ in terms of the Dirac delta ``function''
$\delta(\chi)$.  We shall call a differential form $F$ on $U^+\cup U^-$ {\it
regularly discontinuous} if the restrictions $F^\pm=F|_\upms$ are smooth and
the (1-sided) limits $F^\pm\AtSigma=\lim_{t\to\Sigma^\pm} F$ exist.  It
follows that
$$
d\bff=\bfy^{+} dF^+ + \bfy^{-} dF^- + \d\wedge [F] \Eqno
\label\FDist
$$
where $[F]:=\fp-\fm$ is the discontinuity in $F$.  If we now postulate the
distributional field equation
$$
d\bff=0\Eqno
$$\label\DEq
then we obtain both the original field equations
$$
d F^\pm =0 \Eqno
$$
and the boundary condition
$$
\d\wedge[F]=0 \Eqno
$$\label\DMatch
This formalism is valid irrespective of whether the metric signature changes
at $\Sigma$.  An example with constant signature is given in the Appendix.

Consider now the manifold $M={\Bbb R}$ with signature-changing metric
$$
ds^2 = t \, dt^2
\Eqno$$\label\Metric
Away from $\{t=0\}$, the Hodge dual operator associated with this metric is
given by
$$
\eqalign{*1 &= \smh\,dt\cr{*}\smh\,dt &= \sgn(t)}\Eqno
$$\label\Hodge
The massless wave equation for a 0-form $\Phi$ on a region of $M$ where the
metric is non-degenerate  may be written
$$
dF=0 \Eqno
$$ \label\FWave
where $F={*}d\Phi$ in terms of the Hodge map $*$ defined by the metric.
Setting $F^\pm={*}d\Phi^\pm$ on $\upm:=\{\pm t>0\}$, where
$\Phi^\pm=\Phi|_\upms$, we thus require that \FWave\ be satisfied on $\upm$,
i.e.\ that $dF^\pm=0$.  We now seek distributional solutions to \DEq\ where
$\bff$ is defined as above.  In order for this to make sense, we admit only
solutions such that the tensor $F$, defined for $t\ne0$ by
$F|_\upms=F^\pm={*}d\Phi^\pm$, is regularly discontinuous at $\s$ so that
$[F]=[*d\Phi]$ is well-defined.
\Footnote{In \HaywardII, Hayward argues that, in the presence of signature
change, $*$ must be interpreted as a distribution.  He further argues that
$d\Phi$ must be $C^{\infty}$ so that $*$ can act on it.  There is no need for
such a requirement; it is only necessary that $[*d\Phi]$ be well-defined.
Furthermore, if $T^\pm$ are tensors on $\upm$ such that the tensor $W$, defined
by $W|_\upms={*}T^\pm$, is regularly discontinuous at $\Sigma$, then a natural
Hodge dual operator $\hat{*}$ can be defined via
$$\hat{*}\left( \bfy^+ T^+ + \bfy^- T^- \right)
  = \bfy^+ {*}T^+ + \bfy^- {*}T^-$$
where $*$ refers to the Hodge dual operators on $\upm$ as appropriate.  If we
now impose the natural condition that $\Phi$ be continuous, then
$$\bff = \bfy^+ {*}d\Phi^+ + \bfy^- {*}d\Phi^- \equiv \hat{*}d{\bmit\Phi}$$
where ${\bmit\Phi} = \bfy^+\Phi^+ + \bfy^-\Phi^-$, so that in this case \DEq\
takes the usual form of the wave equation.}
Using \Hodge\ we see that
$$
*d\Phi^\pm = \sgn(t){\p_t\Phi^\pm\over\smh} \Eqno
$$
If $\p_t\Phi^\pm/\smh$ is bounded as $t{\to}0^\pm$ then $*d\Phi$ is regularly
discontinuous at $\s$ and
$$
[*d\Phi] = \Lim0+{\p_t\Phi^+\over\smh} + \Lim0-{\p_t\Phi^-\over\smh} \Eqno
$$ \label\Junction
Then \DMatch\ implies
$$
dt \wedge [*d\Phi] = 0 \Eqno
$$
which for this simple example is equivalent to
$$
[*d\Phi] = 0 \Eqno
$$\label\Bnd
since $M$ is 1-dimensional.

In summary, the distributional field equation \DEq\ requires satisfaction not
only of the wave equations on each side, namely
$$
d{*}d\Phi|_\upms=0 \Eqno
$$ \label\DWave
but also of the boundary conditions \Bnd.
Introducing new ``time'' parameters $\tau$ for $t<0$ and $\sigma$ for $t>0$ by
$$
\eqalign{\tau &= \int_0^t \sqrt{-t} \,dt\cr
  \sigma &= \int_0^t \sqrt{t} \,dt}\Eqno
$$
allows us to rewrite \Bnd\ as
$$
\Phi,_\sigma\AtSigma = -\Phi,_\tau\AtSigma \Eqno
$$ \label\TauMatch
(Changing the relative orientations of $\upm$ amounts to inserting a factor
of $\sgn(t)$ into both of equations \Hodge, resulting in \TauMatch\ without the
minus sign; this gives \Match\ for the 1-dimensional example discussed
previously.)

\Section{GENESIS OF EUCLIDEAN METHODS IN QUANTUM THEORY}

It is of interest to recall the genesis of the use of Euclidean methods in
{\it quantum theory}.  A convenient way to determine the spectrum $\{E_n\}$ of
the quantum Hamiltonian $H$ and its associated energy eigenfunctions
$|n\rangle$ in the position representation $|x\rangle$ is to use the
generating function
$$
\langle x_f | e^{-HT} | \, x_i\rangle =
	\sum_n e^{-E_n T} \langle x_f |n\rangle \langle n|\, x_i\rangle
\Eqno$$
where $T$ is a real parameter.  The leading term in this expression for large
$T$ gives the energy and lowest energy wavefunction.  The left hand side of
this equation may be represented as a path integral over all trajectories
$x(\tau)$ satisfying $x(0)=x_i$ and $x(T)=x_f$
$$
\langle x_f | e^{-HT} |\, x_i\rangle={\cal A}_0\int e^{S_T} \, [dx]
\Eqno$$
where in terms  of the classical potential $V$
$$
S_T=\int_0^T \left( (dx/d\tau)^2/2 +V(x(\tau)) \right) \, d\tau
\Eqno$$
This representation has the advantage that it can be approximated in the
semi-classical limit where the integral is dominated by the stationary paths
${\bf x}$ of $S_T$.

For potentials with a double-well shape the stationary paths ${\bf x}$ of
$S_T$ are the instanton solutions of the classical Euclidean equations of
motion. By considering the fluctuations about the stationary points one can
derive the standard transmission amplitude for a particle to tunnel through a
potential barrier. In this case the stationary path ${\bf x}$ is known as the
``bounce''. For such a solution the modulus of the transmission amplitude
through a potential barrier of width $x_2-x_1$ is given approximately by
$$
\exp\Big( -S_\infty({\bf x}) \Big)
  \equiv\exp\left( -\int_{x_1}^{x_2}\sqrt{2V(x)}\,dx \right)
\Eqno$$
in full accord with the WKB approximation. It should be noted that the bounce
is the zero ``energy'' Euclidean solution corresponding to vanishing ``kinetic
energy'' at $T=\infty$ and the width of the Euclidean domain is determined by
the potential. One might proceed in the semi-classical mode of thought and
pretend that the classical particle materialises after barrier penetration
from a Euclidean domain and evolves according to Newton's laws of motion
starting with zero momentum.  However no-one would demand that quantum
mechanics requires that all classical solutions have such a restrictive
initial condition.

The above methodology exploits the analyticity of the standard Lagrangian for
Newtonian mechanics in the evolution parameter to Wick rotate to {\it
imaginary} time and dominate the continued action with {\it particular} real
solutions to the continued equations of motion. With obvious caveats this
philosophy has been generalised to field systems that possess the required
analyticity.

When one deals with curved space-time metrics the above Wick rotation is not
available and the lack of analyticity of the Riemannian volume element
requires somewhat ad hoc continuation methods.  For problems in quantum
cosmology where the trajectories are 3-geometries, the insistence on real
tunneling solutions across degenerate geometries that dominate the action
gives rise to a behaviour analogous to that of the zero energy bounce
solutions in particle mechanics.

The important point to stress is that in all these cases it is {\it quantum
amplitudes} that are being defined by finite action contributions to the path
integral. The properties of particular Euclidean configurations that enter
into such a description are not shared by the general classical configurations
that can be patched across degeneracy hypersurfaces to Lorentzian signature
domains. It is our purpose to consider the so called theory of ``classical
signature change'', i.e. to treat field theories on a manifold with a
degenerate metric tensor field in a coherent manner without recourse to any
path integral technology.  The quantisation of such configurations is a
separate issue and deserves independent attention.

\Section{CONCLUSION}

We have shown that Hayward's criticisms of our work are untenable.  Our field
equations make sense and our solutions do not suffer from a lack of
``predictability''.  The question of what conditions are the most appropriate
to describe the physics arising from the propagation of classical and quantum
fields in the presence of a background which exhibits signature change can
ultimately only be resolved by observation.  Until then, the implications of
all reasonable, internally consistent approaches should be investigated.

\bigskip\leftline
{\bf APPENDIX: Discontinuities in electromagnetic fields}\nobreak

The first two of Maxwell's equations for a stationary electromagnetic field
are
$$dE = 0 = d{*}B$$
where $*$ is now the Hodge dual operator on Euclidean ${\Bbb R}^3$.  At a
boundary $\{\chi=0\}$ in ${\Bbb R}^3$ we write each of the fields $F={*}B,E$
as in \Dist.  Postulating the distributional versions \DEq\ of Maxwell's
equations leads to Maxwell's equations on each side of the boundary together
with the boundary conditions
$$
d\chi\wedge [F] =0
$$
This gives the standard boundary conditions:
$$
d\chi\wedge [E] =0
$$
$$
d\chi\wedge [*B] =0
$$
The remaining two equations are
$$d{*D} = {*}\rho \quad~~ dH = {*}J$$
where $\rho$ is the electric charge density and $J$ the electric current.  A
similar argument including sources shows that if $J$ and $\rho$ contain
distributional components $J^\Delta \d$ and $\rho^\Delta \d$, respectively,
then they must be related to the discontinuities of $H$ and $D$ by the
standard boundary conditions:
$$
d\chi\wedge [H] ={*}J^\Delta\wedge d\chi
$$
$$
d\chi\wedge [*D] ={*}\rho^\Delta\wedge d\chi
$$
in terms of surface distributions of charge and current.

\bigskip\leftline{\bf ACKNOWLEDGMENTS}\nobreak

We gratefully acknowledge useful discussions with David Hartley and Philip
Tuckey.  TD and CAM would like to thank the School of Physics \& Chemistry at
Lancaster University for kind hospitality during their sabbatical visit.  This
work was partially supported by NSF Grant PHY-9208494 (CAM \& TD), a Fulbright
Grant (CAM), and the Human Capital and Mobility Programme of the European
Union (RWT).

\References

\bye